# Modeling therapy sequence for advanced cancer: A microsimulation approach leveraging Electronic Health Record data

Elizabeth A. Handorf, J. Robert Beck, Daniel M. Geynisman

**Abstract**

**Purpose**

Many patients with advanced cancers undergo multiple lines of treatment. Optimal therapy sequence, i.e., what therapies are given and in what order, is often unknown. We develop methods for estimating quality-adjusted outcomes and cost-effectiveness of therapy sequences, informed by patient-level longitudinal data from Electronic Health Records (EHRs).

**Methods**

We develop microsimulation models with a discrete-time health-state transition framework and propose two methods: one using multi-state models to estimate transition probabilities, and one using observed patient trajectories through the health states. We use bootstrap resampling to estimate standard errors. We create synthetic EHR-like datasets to evaluate these methods where within-patient transition times depend on covariates and a copula generator, and compare with Markov cohort models. We demonstrate these methods in two treatment sequences for advanced bladder cancer (cisplatin or carboplatin-based therapy followed by immunotherapy), incorporating external information on costs, utilities, and expected adverse event.

**Results**

Both methods produced well-calibrated overall survivals, although the trajectory approach was often superior. The multi-state model approach generated lower standard errors but was biased when compared to known results from the synthetic datasets. The observed trajectory approach mostly produced confidence intervals that covered known values. In the bladder cancer example, both methods result in a Net Monetary Benefit (NMB)>0 for the cisplatin-based treatment sequence with a willingness to pay of $100,000 per quality-adjusted life year. For the multi-state model approach, NMB = $11,185 (95%CI $7,477, $14,893), trajectory approach NMB= $12,572, (95% CI $3,532, $21,611).

**Conclusions**

Both microsimulation methods produce well-calibrated results and offer superior performance to a homogeneous Markov cohort approach when studying therapy sequence.  Where available, patient level EHR-based data should be considered to inform cost-effectiveness models.

1. Background and motivation

In modern oncology practice, new anti-cancer therapies frequently offer improved survival and tolerability with hundreds of new drugs approved since 2000.[1] Therefore, more patients are living longer with advanced cancers.[2] Often such patients undergo multiple lines of therapies, switching treatments as needed. For many patients and scenarios, the clinically optimal order of therapies is not clear and rarely studied by prospective randomized trials, with new treatment paradigms continuing to evolve. Therefore, it is critical to evaluate different sequences of therapies, considering the whole planned course of treatment, not just treatment decisions at a single timepoint. Here, we define therapy sequence to be either the order in which therapies are given or the use of agents at specific times during a multi-line treatment plan. Because of the availability of longitudinal data from Electronic Health Records (EHRs), it is now possible to study therapy sequence questions in a real-world setting.[3] In this paper, we develop modeling approaches for cost-effectiveness analysis in a heterogeneous clinical cohort including within-patient effects.

The high price of many anti-cancer drugs makes cost-effectiveness analysis very important. Although generally under-studied, the cost-effectiveness of various therapy sequences has been examined in several advanced cancers, including BRAF wild-type melanoma[4], EGFR mutated non-small-cell lung cancer,[5,6] HER2+ breast cancer,[7,8] and KRAS wild-type colorectal cancer.[9] These cost-effectiveness analyses often use state-based Markov cohort models.[4,7–9] This comes with two main limitations: the use of a single homogeneous cohort and the memoryless assumptions of the Markov model. These limitations may lead to unrealistic simplifying assumptions. Recent studies of therapy sequence have used microsimulation models.[10] Microsimulations allow outcomes to depend on past states, and readily incorporate patient characteristics, giving a higher-fidelity approximation of the true clinical scenario. [11–13] Previously, we presented a study of therapy sequence in metastatic castrate-sensitive prostate cancer using a microsimulation approach with a homogenous cohort, with survival based on published trial results. This approach achieved calibration by using a penalty function applied to treatments after the first line of therapy, when a simulated patient had poor progression-free survival in the first line.[14]

As a motivating example, we consider the clinical question of carboplatin/gemcitabine (carbo/gem) vs cisplatin/gemcitabine (cis/gem) first-line therapy followed by immunotherapy (IO) for advanced urothelial carcinoma. Cisplatin provides superior survival to carboplatin but comes with generally more toxicities than carboplatin that reduce quality of life.[15] Recent work has shown that the survival benefit of cisplatin may be smaller now that IO is available to treat patients who progress on first-line platinum-based therapy,[3] making toxicity considerations more important. Therefore, obtaining estimates of quality-adjusted life years (QALYs) is of clinical interest. Throughout this paper, we utilize a bladder cancer dataset extracted from the nationwide Flatiron Health EHR-derived database.[16,17] Similar to other EHR-based datasets, it does not contain cost or quality of life measures, but it provides a rich source of data on treatments, outcomes, and patient-level covariates.

In this work, we propose to use real-world patient data based on EHRs to better model within-patient dependence across multiple lines of therapy. This can be implemented via a microsimulation approach, where the trajectory of each patient through a health-state model is governed based on their covariates and prior transition times. We show how this can be accomplished in two ways: by using multi-state model fitted transition probabilities, and by using individual patient trajectories with bootstrap resampling. We evaluate this approach in simulated datasets with complex correlation structures and demonstrate its use in a cost-effectiveness analysis of therapy sequence for advanced or metastatic bladder cancer.

## 2. Methods

We use the basic model structure shown in Figure 1. Patients with advanced or metastatic (incurable) disease begin on some "Line 1" (L1) therapy. After progression, they may begin "Line 2" (L2) therapy. Some patients may not receive for second-line therapy; these we assume enter a health state called "Extensive Disease" (ED). Patients may die in any state in the model. This framework could be augmented to contain extra health states (see section 4 for an extension with adverse event states), or additional potential lines of therapy. As is standard for health-state models, patients accrue costs and quality-adjusted survival time in each state. Here, we use a discrete time model, with patients transitioning between model states at the start of each cycle.

**Figure 1: Model structure**

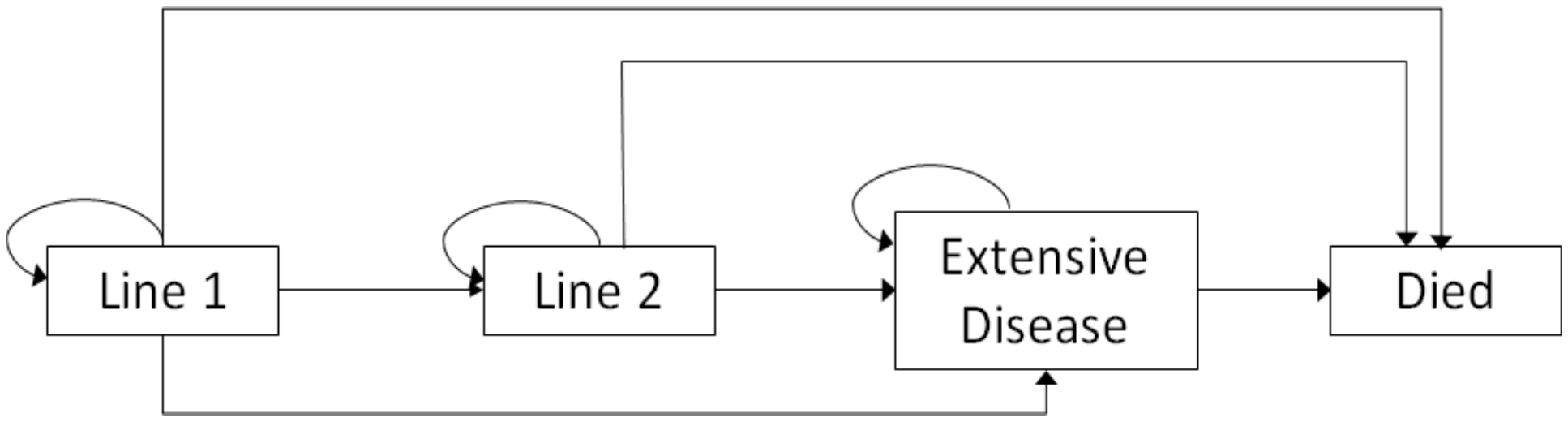

We use a microsimulation framework for this analysis, wherein we simulate individual patient trajectories through the model. Code use to implement our methods is available at https://github.com/BethHandorf/Therapy_sequence_EHR .

### 2.1 Multi-state modeling

Multi-state models are a useful class of models for frameworks containing multiple health states, and competing or intermediate events.[18] The model framework shown in Figure 1 is readily accommodated by multi-state models. In this paper we use multi state models, which assume that the hazard functions defining each transition from state *g* to state *h* can be represented by Cox proportional hazards models,

$$\lambda_{gh}(t) = \lambda_{gh0}(t) exp\left(\alpha_{gh} Z + \beta_{gh} X\right).$$

Where $\lambda_{gh0}(t)$ is the baseline hazards, Z is the treatment sequence, $\alpha_{gh}$ is the effect of the treatment sequence on transition *g,h*. We assume that patient covariates, X, impact transition probabilities, so let $\beta_{gh}$ be the covariate effects on transition *g,h*.

Based on these equations, a given treatment sequence, and an individual's covariate values, we can compute subject-specific cumulative transition hazards, which are then used to compute subject and time-specific transition probabilities. These methods have been described in detail by Putter et al.[18], and software is available through the R package mstate.[19]

For efficiency, we recommend pre-calculating the subject-specific cumulative transition hazards. These can then be used to estimate the transition probabilities, (given subject *i*, current state *s*, and time *t*), during each iteration of the microsimulation model. For further efficiency, we have found in some cases that transition probabilities can be estimated once per state, during the first cycle after a transition occurs, and using that probability for subsequent states (i.e. making an exponential assumption about subsequent transition probabilities).

The microsimulation is run as follows: A patient from the full study cohort of N patients, with covariate patterns $X_i$ enters into the L1 state. We then use the fitted multi-state model determine the transition probabilities for the patient in cycle 1 given $X_i$, assuming they were treated with treatment sequence 1. We randomly draw the patient's state for the start of the next cycle based on these transition probabilities (which include a probability of remaining in the current state). If the state changes, a transition has been observed. This process is repeated for T total cycles. During each cycle, the patient accrues costs and effects (here we use quality adjusted life years, or QALYs). We run this simulation for separately for each patient in the cohort (microsimulation).[13] We then repeat this process, assuming each patient is treated with sequence 2. When aggregated, this provides us with expected costs and effects for the full cohort if they were treated with sequence 1, and for sequence 2. We refer to this as our *mstate* approach.

An important part of model validation is to assess calibration. We recommend comparing the overall survival curve that results from the aggregated transitions in the microsimulation model to the overall survival curve from the EHR data. To account for the non-randomized nature of the data, one can estimate a Kaplan-Meier curve with Inverse Probability of Treatment Weighting using propensity scores.[20] When using Average Treatment Effect weights, this IPTW curve is directly comparable to the results of the microsimulation model.[21] Differences between the two curves suggest lack of fit. This can be formally assessed using the Kruskal-Wallis test.

Finally, we need to obtain estimates of uncertainty. In a microsimulation with a homogenous cohort, one can run the microsimulation many times with different seeds to obtain estimates of variability. In our case, however, we need to account for simulation-based and population-based variability. We therefore use a bootstrapping approach, where we re-sample the cohort with replacement and re-fit the microsimulation. We use a different randomization seed for each bootstrap replicate; therefore, our bootstrap estimates of standard error will account for both sources of variability.

## 2.2 Observed patient trajectories

In the setting of EHR data, we have a unique opportunity to use directly observed patient transitions between states instead of model-based predicted probabilities. If data were fully observed and subjects were balanced (i.e. through randomization), we could simply aggregate the time spent in each state. Nevertheless, in real-world applications, data are likely to be censored for some proportion of the population, and patient balance is unlikely, as confounders may affect treatment decisions and survival outcomes. Finally, it is important to obtain estimates of variability. We propose the following strategy to use the observed patient data while overcoming these issues. We will refer to this as our *trajectory* approach.

First, we consider censoring. We propose to use the observed patient transition times until censoring occurs. This can be simply done in the microsimulation model by setting all of a patient's transition probabilities to 0 unless a particular transition is observed to occur; if so, we set that probability to 1. After censoring occurs, we can use the estimated transition probabilities obtained using the multi-state model, as described above.

Second, we need to account for imbalance in patient characteristics between treatment groups. In the model above we can predict outcomes for all patients in the cohort, but when using actual trajectories, we can only observe trajectories from those who receive a particular treatment. We propose to use propensity-score based inverse probability of treatment weighting to account for this imbalance. Propensity score weighing has been successfully used to balance observed characteristics in observational studies of treatment alternatives, and this technique has been described in detail in the statistical literature.[20] Briefly, the propensity score is defined as the probability of treatment given covariates, and can be estimated by logistic regression or any appropriate prediction model.[22] Here we use average treatment effect weights.[21] Patients who received a treatment that they have a low probability of getting are upweighted. This has the result of creating a pseudo-population where covariates are balanced. We center the weights at 1 to maintain the original population size. We apply the propensity score weights after the microsimulation model has run. We re-weight each patient's costs and effects, so that when costs and effects are summed in the full treatment groups, their results are directly comparable.

Finally, we can account for variability using a bootstrap approach, similar to that described above. Within each treatment arm, we resample with replacement, and then run the microsimulation with varying seeds. This bootstrap sampling can be used to estimate the bootstrap standard errors.

## 3. Application to synthetic EHR datasets

We created a synthetic EHR dataset to assess these methods under a variety of conditions. To avoid confusion with the terminology "microsimulation", we term this data set as "synthetic" instead of "simulated". To assess how well this approach works in realistic datasets, we generated covariate and outcome data by drawing from distributions using parameters estimated from the Flatiron Health bladder cancer dataset. Additional detail about the bladder cohort is discussed in Section 4.

### 3.1 Synthetic dataset creation

We first created a cohort of 2,000 subjects. We assessed the marginal distributions of binary, ordinal, and continuous covariates in the Flatiron Health bladder dataset, fitting parametric models in the observed dataset for each covariate. For continuous variables (e.g. age, lab values), we used the R package fitdistrplus to obtain parameter estimates and chose functional forms which best matched the observed distributions We then estimated the correlation structure of the covariates by Spearman's correlation. We then drew from the multivariate normal distribution with standard normal marginals and the estimated correlation matrix found in the data. Covariates were then transformed to the respective fitted marginal distribution via a probability integral transformation. Distributions for continuous covariates were as follows: Age used a transformed Beta distribution (to account for age cutoffs in the dataset); glomerlular filtration rate (GFR) and hemoglobin used a Gamma; Neutrophil Lymphocyte Ratio (NLR) used a Weibull. We fit a logistic regression model predicting treatment sequence in the Flatiron Health bladder dataset. We then used this model to predict treatment probabilities in the synthetic dataset and drew treatment status from the binomial distribution. Model parameters are given in the supplemental tables.

We next drew transition times from survival regression models, with additional correlations between outcomes within-patients implemented via a copula model. We anticipate substantial within-patient correlation in longitudinal datasets. Some of the within-patient correlation can be attributed to observed patient characteristics and is readily modeled by covariate effects in the survival models. We anticipate that in some cases, there may be additional correlation due to unobservable factors (e.g. tumor aggressiveness). We can introduce additional correlation via copula models, which allow us to draw outcomes from arbitrary marginal outcome distributions, with correlations imposed via a generator function. Copulas have been used previously to simulate multi-state model transitions.[23] We chose Clayton's copula as our base case. This produces data where correlations tend to be larger for larger outcome values. We also used Gaussian copulas, where correlations are constant across the range of outcome values, and T copulas, where correlations are larger at the extremes.[24] These copulas used compound symmetric or unstructured correlations.

We obtained parameter estimates for the marginal survival models using the Flatiron Health bladder dataset. We then used these parameters to randomly draw outcomes from three survival models: Weibull (base case), log-normal, and log-logistic. For the synthetic data, we simultaneously drew all potential transition times for the following events: progression from L1, death from L1, progression from L2, death from L2, and death from extensive disease. The event with the smallest time in each state is therefore the event which occurred. Transitions to L2 or ED after progression on L1 were drawn separately, conditional on covariates but not transition times. Censoring times were drawn from the uniform with a range of 6 to 72 months. We created two additional datasets: one if the full cohort received treatment sequence 1 (seq1), and one if everyone received treatment sequence 2 (seq2), with censoring occurring at 60 months. These datasets allow us to determine true outcomes for comparison.

We generated synthetic datasets for 9 different scenarios. We fit Weibull models with 6 different copula structures: Independence, Clayton, Gaussian compound symmetric, Gaussian unstructured, T compound symmetric, T unstructured. (See technical appendix.) We also assessed the effects of a small sample size (500 patients) with Weibull models and Claytons copula. Additionally, we assessed log-logistic and log-normal survival models with Clayton's copula.

For each of the 9 datasets, we then used the two microsimulation methods described above in section 2. For each health state, we included costs and utilities (see Section 4). Note that we did not include adverse events in the analysis of the synthetic datasets. The microsimulations were run for a period of 60 months (5 years). We obtained estimates of bootstrap standard errors using 100 replicates. For comparison, we also fit a cohort-based Markov model.[25] Here, we estimated transition probabilities using cause-specific Weibull survival models, using inverse probability of treatment weighting to account for differences in covariates between the two treatment sequences.

Finally, we obtained the "true" values by creating two synthetic datasets, assigning all patients to each treatment sequence of interest, with no censoring until the end of the study period. Calculating the total costs and effects over the study period in the synthetic cohorts with full data provides us with a standard for comparison to assess bias and coverage. We then compared the mstate and trajectory microsimulation results to the true values for costs, effects (Quality Adjusted Life Years (QALYs)), and Net Monetary Benefit (NMB) with a willingness to pay of $100,000.

## 3.2 Results

Generally, both modeling strategies did quite well at producing overall survival curves with good fit, however, the trajectory method was often closer to the IPTW curves than the mstate method. In all but one case, the sum of squared errors between the two curves was smaller when using the trajectory approach. See Table 1 and Figure 2A and 2B for a comparison of the trajectory and mstate results. In all scenarios, the trajectory approach had adequate fit (Kruskal-Wallis $P>0.05$). For the mstate approach, only one scenario formally failed a test for lack-of-fit, treatment sequence 2 for the log-normal Clayton scenario (see Table 1 and Figure 2C). The mstate model in the log-logistic scenario was also not as well calibrated, but this was not statistically significant. We see improved calibration over a traditional Markov model approach (Figure 2D).

**Figure 2: calibration curves of selected models**

A. Weibull model, Claytons copula, trajectory approach

Sequence 1 Sequence 2

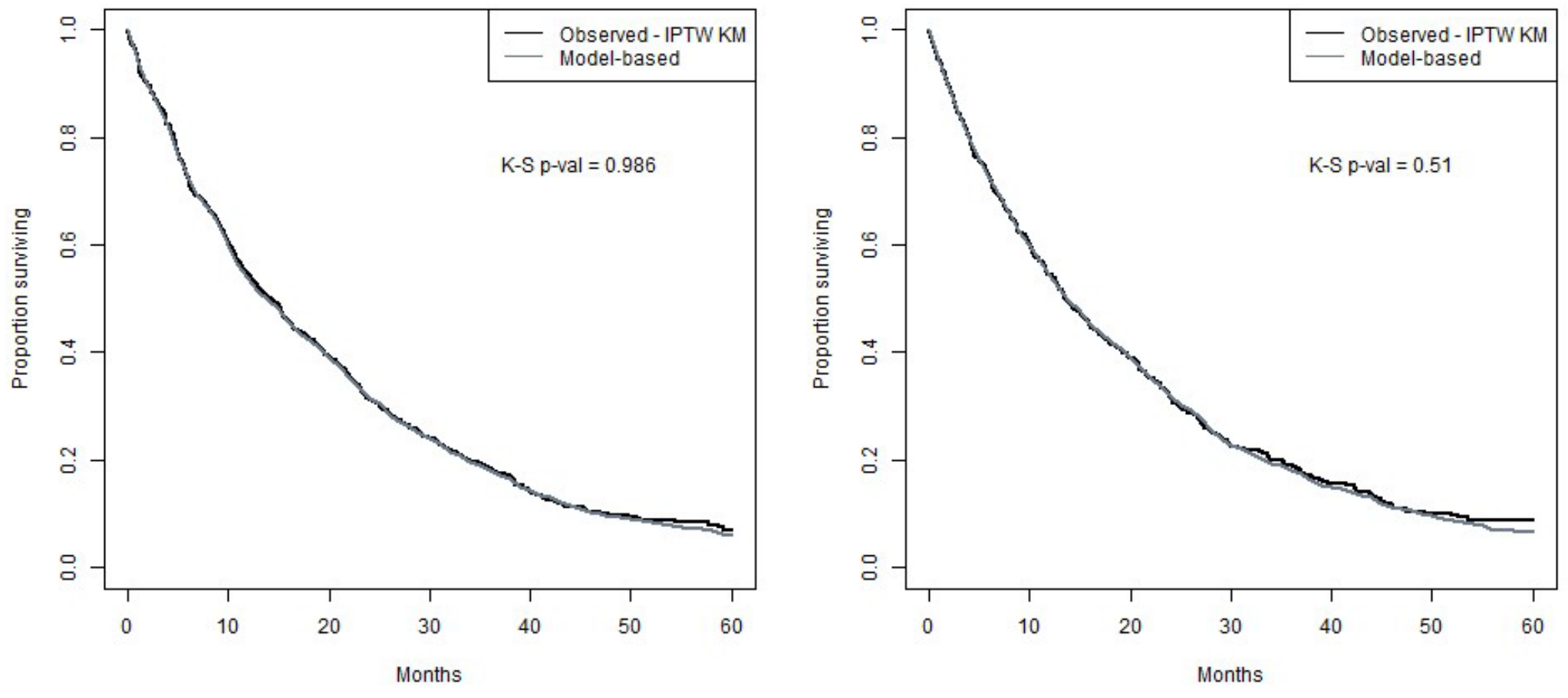

B. Weibull model, Clayton's copula, mstate approach

Sequence 1 Sequence 2

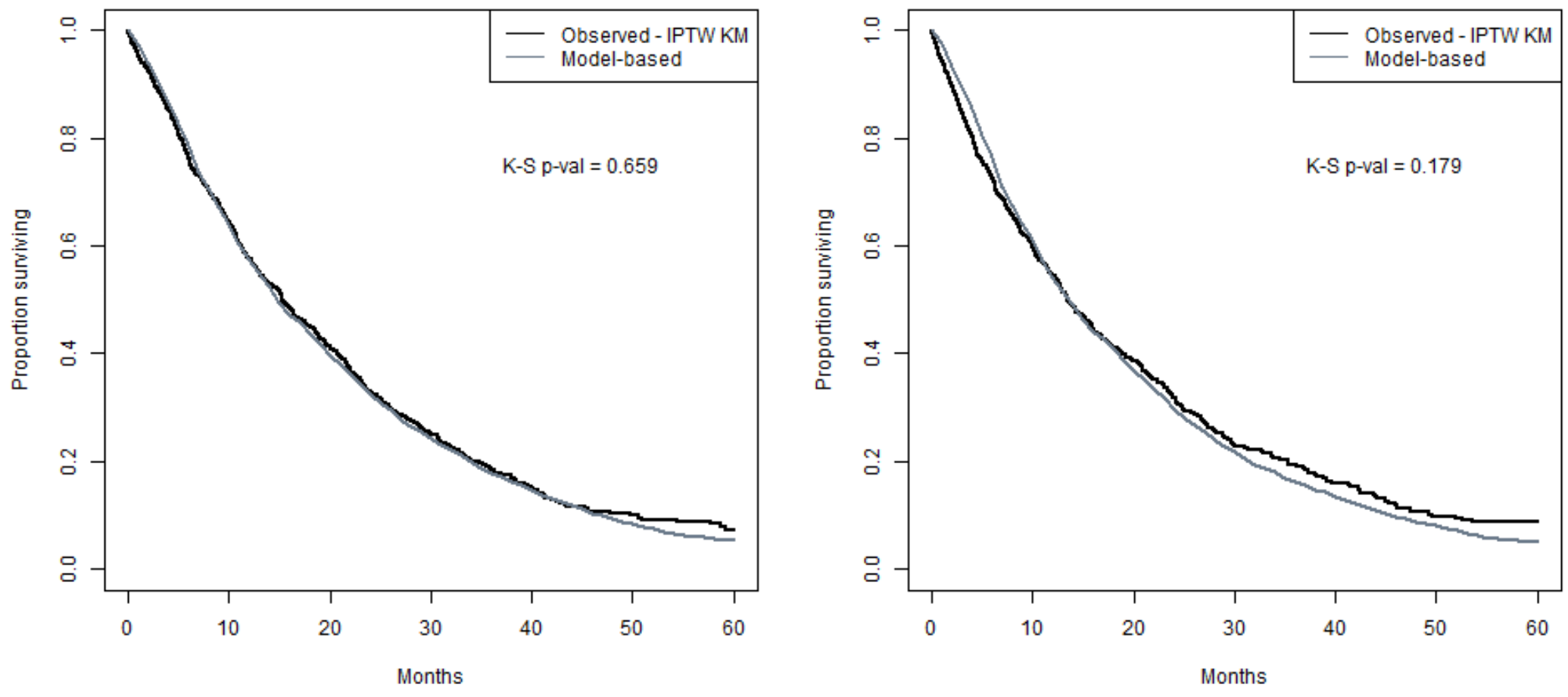

C. Log-normal model, Clayton's copula, mstate approach

Sequence 1 Sequence 2

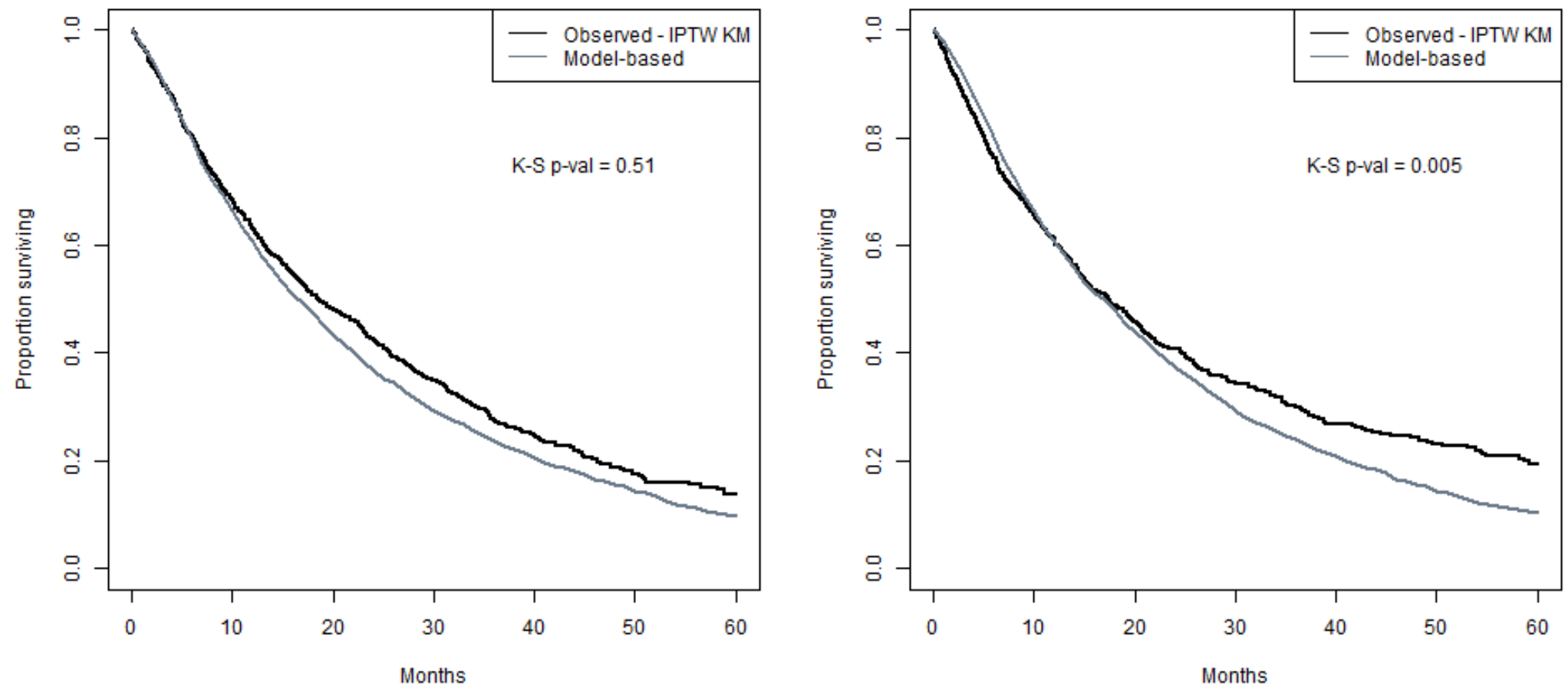

D. Weibull model, Clayton’s copula, standard Markov cohort models

Sequence 1 Sequence 2

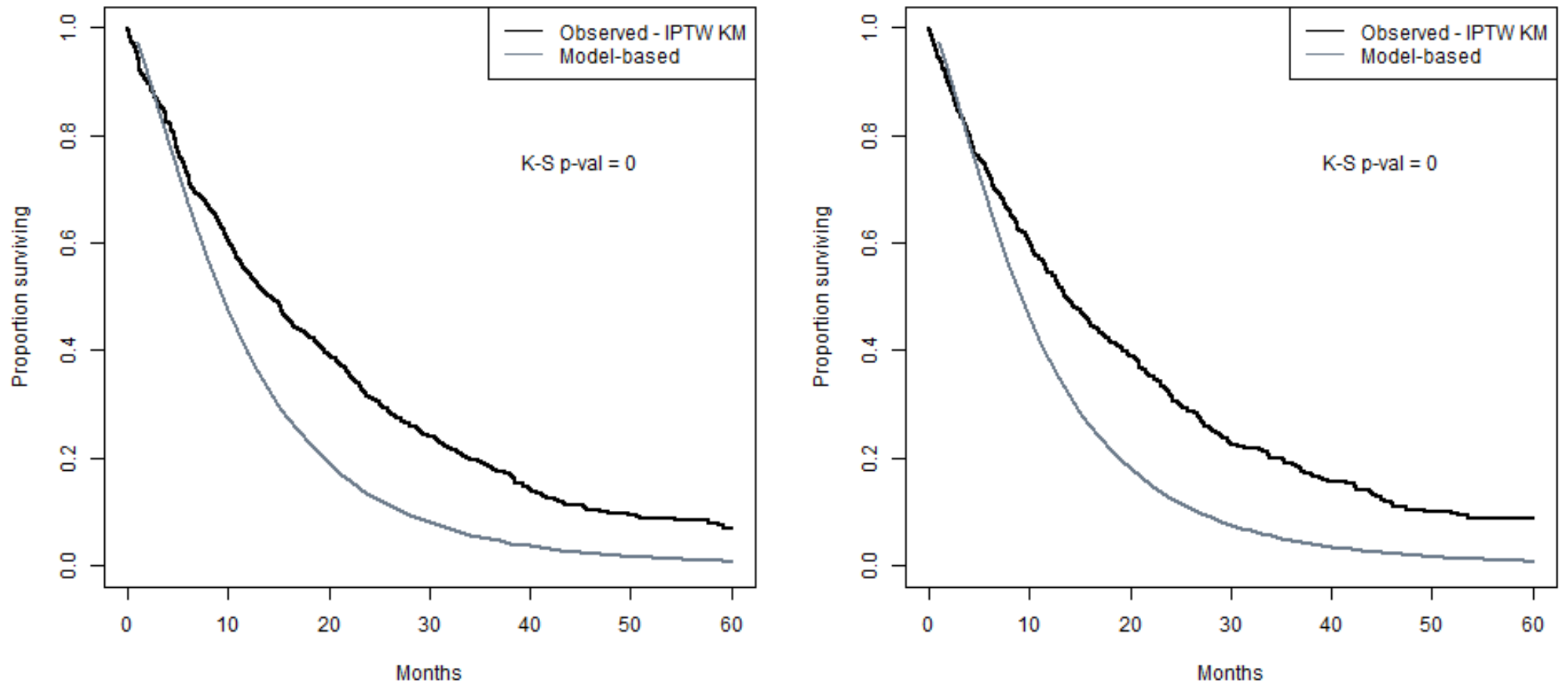

**Table 1: Costs and effects estimated from microsimulation models, with sum of sum-of-squared-errors and goodness-of-fit p-values comparing the microsimulation-based and weighted observed the overall survival curves**

| | | Mstate Model-based | | | | Observed trajectories | | | |
|---|---|---|---|---|---|---|---|---|---|
| | | Cost | Effect | Sum Sq Err | K-S p-val | Cost | Effect | Sum Sq Err | K-S p-val |
| Independence | Seq2 | 111122 | 0.91 | 2.364 | 0.660 | 110418 | 0.89 | 0.055 | 0.119 |
| | Seq1 | 120588 | 1.00 | 0.027 | 0.990 | 127873 | 1.06 | 0.115 | 0.999 |
| Clayton | Seq2 | 117241 | 0.98 | 0.515 | 0.180 | 129328 | 1.00 | 0.056 | 0.510 |
| | Seq1 | 123641 | 1.04 | 0.011 | 0.660 | 123718 | 1.02 | 0.001 | 0.990 |
| Gauss compound symmetric | Seq2 | 112099 | 0.94 | 0.048 | 0.990 | 115977 | 0.92 | 0.046 | 0.990 |
| | Seq1 | 122141 | 1.04 | 0.013 | 0.660 | 121481 | 1.01 | 0.001 | 0.990 |
| Gauss unstructured | Seq2 | 108877 | 0.90 | 0.597 | 0.990 | 116863 | 0.92 | 0.042 | 0.980 |
| | Seq1 | 119804 | 1.01 | 0.015 | 0.370 | 120356 | 1.00 | 0.004 | 0.930 |
| T compound symmetric | Seq2 | 113501 | 0.95 | 0.175 | 0.810 | 122033 | 0.96 | 0.025 | 0.990 |
| | Seq1 | 119950 | 1.02 | 0.019 | 0.660 | 127220 | 1.04 | 0.003 | 0.990 |
| T unstructured | Seq2 | 114509 | 0.94 | 0.097 | 0.510 | 121286 | 0.96 | 0.077 | 0.990 |
| | Seq1 | 124550 | 1.03 | 0.015 | 0.660 | 128780 | 1.05 | 0.002 | 0.990 |
| Small sample size (Clayton) | Seq2 | 120837 | 0.99 | 0.137 | 0.650 | 134494 | 1.07 | 0.479 | 0.370 |
| | Seq1 | 113730 | 0.94 | 0.327 | 0.790 | 124287 | 1.00 | 0.010 | 0.370 |
| Log-logistic (Clayton) | Seq2 | 97098 | 0.90 | 0.049 | 0.120 | 94748 | 0.84 | 0.048 | 0.510 |
| | Seq1 | 97046 | 0.92 | 0.328 | 0.510 | 89020 | 0.81 | 0.004 | 0.180 |
| Log-normal (Clayton) | Seq2 | 136942 | 1.14 | 6.054 | 0.005 | 155294 | 1.21 | 0.386 | 0.660 |
| | Seq1 | 135885 | 1.14 | 0.092 | 0.510 | 143721 | 1.17 | 0.004 | 0.990 |

As this is fully synthetic data, we can also compare both methods to the “true” values which result from synthetic data where all patients are given each treatment sequence, and no censoring occurs.  We can calculate costs and QALYs based on the observed time spent in each health state.  Table 2 shows a comparison of each model results to the true target values.  We find that both methods tended to underestimate the difference in effects, particularly when using Clayton’s copula.  The trajectory approach performed best under the Gaussian and T-copula scenarios.  The mstate approach did not cover the true values in most of the scenarios. Concerningly, even with a Weibull model (which is a proportional hazards model), and with independence between states (conditional on covariates), the mstate model still exhibited bias, with confidence intervals which did not cover the true value.  Results in both approaches were particularly problematic for the small sample size case, log-logistic, and log-normal models, where cost and effects differences were in the wrong direction.

Generally, the trajectory approach was more accurate.  This method tended to have wider confidence intervals, as well as more accurate estimates. The confidence intervals for the trajectory approach often covered the true value, which was not the case for the mstate approach. Therefore, the trajectory approach should be preferred here.  However, we note that both methods offered a large improvement over the traditional homogenous Markov cohort approach.  In our synthetic data, this approach demonstrated markedly poor fit, underestimating both costs and effects (See Figure 2D).  Nevertheless, these results show that even good model calibration does not guarantee unbiased results.

**Table 2: Differences in costs, differences in effects, and NMB estimated from microsimulation models, compared with actual values observed in synthetic datasets. Gray highlighted cells indicate the confidence intervals do not cover the true values.**

| | | True Value | Mstate Estimate | Mstate LCL | Mstate UCL | Trajectory Estimate | Trajectory LCL | Trajectory UCL |
|---|---|---|---|---|---|---|---|---|
| Independence | Δ Cost | 11860 | 9466 | 7254 | 11678 | 17456 | 6578 | 28333 |
| | Δ Effect | 0.15 | 0.09 | 0.07 | 0.11 | 0.17 | 0.08 | 0.26 |
| | NMB | 3166 | -404 | -1655 | 848 | -856 | -4903 | 3191 |
| Clayton | Δ Cost | 10503 | 6400 | 4292 | 8508 | -5610 | -18208 | 6988 |
| | Δ Effect | 0.15 | 0.07 | 0.05 | 0.08 | 0.01 | -0.08 | 0.11 |
| | NMB | 4212 | 179 | -1044 | 1401 | 7037 | 2376 | 11698 |
| Gauss compound symmetric | Δ Cost | 10856 | 10043 | 7877 | 12208 | 5504 | -4810 | 15818 |
| | Δ Effect | 0.15 | 0.10 | 0.08 | 0.12 | 0.09 | 0.01 | 0.17 |
| | NMB | 3646 | -299 | -1513 | 915 | 3468 | -1074 | 8010 |
| Gauss unstructured | Δ Cost | 11109 | 10928 | 8758 | 13097 | 3493 | -6590 | 13577 |
| | Δ Effect | 0.15 | 0.11 | 0.09 | 0.12 | 0.08 | 0.00 | 0.17 |
| | NMB | 3519 | -345 | -1638 | 947 | 4448 | -64 | 8959 |
| T compound symmetric | Δ Cost | 10911 | 6449 | 4361 | 8536 | 5187 | -6621 | 16994 |
| | Δ Effect | 0.15 | 0.07 | 0.05 | 0.09 | 0.09 | 0.00 | 0.167 |
| | NMB | 4118 | 636 | -570 | 1842 | 3323 | -1805 | 8451 |
| T unstructured | Δ Cost | 10911 | 10041 | 6212 | 13869 | 7494 | -4872 | 19860 |
| | Δ Effect | 0.15 | 0.09 | 0.06 | 0.12 | 0.09 | 0.00 | 0.17 |
| | NMB | 4118 | -1089 | -3058 | 880 | 1215 | -4252 | 6681 |
| Small sample size (Clayton) | Δ Cost | 12181 | -7108 | -10937 | -3279 | -10207 | -31880 | 11467 |
| | Δ Effect | 0.16 | -0.05 | -0.08 | -0.02 | -0.07 | -0.24 | 0.10 |
| | NMB | 3912 | 2262 | 293 | 4231 | 3437 | -4843 | 11718 |
| Log-logistic (Clayton) | Δ Cost | 5624 | -51 | -3880 | 3777 | -5728 | -14976 | 3519 |
| | Δ Effect | 0.08 | 0.02 | -0.01 | 0.05 | -0.03 | -0.11 | 0.05 |
| | NMB | 2766 | 1883 | -86 | 3852 | 2903 | 40 | 5767 |
| Log-normal (Clayton) | Δ Cost | 4250 | -1056 | -3010 | 897 | -11573 | -26148 | 3003 |
| | Δ Effect | 0.08 | 0.00 | -0.02 | 0.01 | -0.05 | -0.16 | 0.06 |
| | NMB | 3335 | 746 | -251 | 1743 | 6681 | 1610 | 11751 |

4. Application to bladder cancer

We next demonstrate our microsimulation approach with the original Flatiron Health EHR-based bladder cancer dataset, using modeling to incorporate costs, quality of life, and adverse events. The Flatiron Health database is longitudinal, comprising de-identified patient-level structured and unstructured data, curated via technology-enabled abstraction.[16,17] We identified patients diagnosed with advanced or metastatic urothelial carcinoma between 2015-2022. The treatment sequences of interest were 1) Line 1 cisplatin/gemcitabine followed by line 2 single agent immunotherapy (atezolizumab, avelumab, pembrolizumab, nivolumab, or durvalumab), or 2) Line 1 carboplatin/gemcitabine followed by line 2 single agent immunotherapy. Patients were included if they received one of these two sequences, or if they started cis/gem or carbo/gem and received no Line 2 treatment during follow-up. Patients were excluded if they had no documented treatments or visits within 90 days of diagnosis or were missing key covariate data. The patient sample is described in Table 3, showing covariates for patients who met all inclusion criteria, and for those who received the full treatment sequences of interest. Real-world progression events were coded via abstraction of clinical notes. [26]

Our sample consisted of 1,804 patients, of whom 911 were treated with the cisplatin-based treatment sequence. After progression on line 1 therapy, 842 received line 2 immunotherapy (of whom 802 had documented line 2 progression), and 568 received no further treatment (extensive disease state). Median follow-up was 39.6 months (reverse Kaplan-Meier method). Unadjusted median progression-free survival was 6.3 months (95% CI 5.9-6.8) for line 1 cis/gem, 4.6 months (95% CI 4.2-5.0) for line 1 carbo/gem, and 2.6 (95% CI 2.5-2.8) months on line 2 therapy. Unadjusted median overall survival was 14.5 months (95% CI 13.1-16.9) for line 1 cis/gem, 9.9 months (95% CI 9.2-10.9) for line 1 carbo/gem, 7.3 months (95% CI 6.2-8.5) after starting Line 2 therapy, and 5.9 months (95% CI 5.2-7.1) in extensive disease.

**Table 3: Cohort characteristics of Flatiron Health bladder cancer dataset.**

| | **Line 1 carbo/gem (N=893)** | **Line 1 cis/gem (N=911)** | **Total (N=1804)** | **p value** | **carbo/gem → IO (N=428)** | **cis/gem → IO (N=414)** | **Total (N=842)** | **p value** |
|---|---|---|---|---|---|---|---|---|
| **Surgery** | 374 (41.9%) | 413 (45.3%) | 787 (43.6%) | 0.139 | 205 (47.9%) | 194 (46.9%) | 399 (47.4%) | 0.763 |
| **Male** | 680 (76.1%) | 652 (71.6%) | 1332 (73.8%) | 0.027 | 324 (75.7%) | 289 (69.8%) | 613 (72.8%) | 0.055 |
| **Race** | | | | | | | | |
| White | 617 (69.1%) | 629 (69.0%) | 1246 (69.1%) | | 302 (70.6%) | 395 (71.0%) | 803 (70.8%) | |
| Black | 42 (4.7%) | 49 (5.4%) | 91 (5.0%) | 0.512 | 20 (4.7%) | 19 (4.6%) | 39 (4.6%) | 0.954 |
| Other/unknown | 234 (26.2%) | 233 (25.6%) | 467 (25.9%) | 0.761 | 106 (24.8%) | 101 (24.4%) | 207 (24.6%) | 0.901 |
| **Hispanic** | 30 (3.4%) | 27 (3.0%) | 57 (3.2%) | 0.631 | 17 (4.0%) | 13 (3.1%) | 30 (3.6%) | 0.515 |
| **Age** | | | | < 0.001 | | | | < 0.001 |
| Mean (SD) | 72.931 (8.617) | 67.982 (8.771) | 70.432 (9.038) | | 72.694 (8.502) | 67.536 (8.865) | 70.158 (9.053) | |
| Range | 31.000 - 85.000 | 32.000 - 85.000 | 31.000 - 85.000 | | 31.000 - 85.000 | 32.000 - 84.000 | 31.000 - 85.000 | |
| **ECOG Value** | | | | < 0.001 | | | | 0.371 |
| N-Miss | 198 | 170 | 368 | | 85 | 71 | 156 | |
| 0 | 253 (36.4%) | 360 (48.6%) | 613 (42.7%) | | 149 (43.4%) | 167 (48.7%) | 316 (46.1%) | |
| 1 | 332 (47.8%) | 313 (42.2%) | 645 (44.9%) | | 161 (46.9%) | 148 (43.1%) | 309 (45.0%) | |
| 2 | 110 (15.8%) | 68 (9.2%) | 178 (12.4%) | | 33 (9.6%) | 28 (8.2%) | 61 (8.9%) | |
| **GFR** | | | | < 0.001 | | | | < 0.001 |
| Mean (SD) | 56.388 (24.878) | 71.388 (23.097) | 63.963 (25.134) | | 55.220 (22.935) | 71.271 (23.631) | 63.112 (24.612) | |
| Range | 5.242 - 180.005 | 5.221 - 197.968 | 5.221 - 197.968 | | 8.756 - 162.994 | 5.221 - 171.389 | 5.221 - 171.389 | |
| **hemoglobin** | | | | < 0.001 | | | | 0.006 |
| Mean (SD) | 11.466 (1.911) | 12.135 (1.847) | 11.804 (1.908) | | 11.742 (1.881) | 12.096 (1.845) | 11.916 (1.871) | |
| Range | 6.700 - 16.800 | 5.900 - 17.400 | 5.900 - 17.400 | | 7.000 - 16.600 | 7.900 - 17.400 | 7.000 - 17.400 | |
| **albumin** | | | | < 0.001 | | | | 0.072 |
| Mean (SD) | 36.954 (5.323) | 38.315 (4.705) | 37.642 (5.065) | | 37.932 (4.324) | 38.500 (4.810) | 38.211 (4.576) | |
| Range | 0.041 - 49.000 | 20.000 - 50.000 | 0.041 - 50.000 | | 21.000 - 49.000 | 20.000 - 48.000 | 20.000 - 49.000 | |
| **CHF known** | 874 (97.9%) | 903 (99.1%) | 1777 (98.5%) | 0.029 | 419 (97.9%) | 411 (99.3%) | 830 (98.6%) | 0.092 |
| **hearing loss known** | 11 (1.2%) | 21 (2.3%) | 32 (1.8%) | 0.084 | 5 (1.2%) | 7 (1.7%) | 12 (1.4%) | 0.522 |
| **neuropathy known** | 31 (3.5%) | 20 (2.2%) | 51 (2.8%) | 0.102 | 14 (3.3%) | 12 (2.9%) | 26 (3.1%) | 0.755 |
| **visceral mets known** | 227 (25.4%) | 165 (18.1%) | 392 (21.7%) | < 0.001 | 110 (25.7%) | 79 (19.1%) | 189 (22.4%) | 0.021 |

We apply the multi-state and trajectory approaches to the clinical dataset, adapting these models to also incorporate adverse events, as shown in Figure 3.  Model inputs for cost, utilities, and probabilities of adverse events were obtained from relevant literature and are shown in Table 4.  For purposes of demonstrating the method, we created one composite adverse event for each treatment. We used the reported overall grade 3/4 adverse event rates in the literature. Many patients experienced more than 1 adverse event (particularly for Line 1 therapy).  We estimated costs for treating this composite adverse event using a weighted sum of the costs of the potential underling events, weighted by the probability of each event. Costs were defined as 2024 dollars; any costs obtained from the literature were updated using the medical consumer price index. Costs and effects were discounted at 3% per year.  We estimated standard errors with 50 bootstrap replicates.

**Figure 3: Model structure applied to bladder cancer dataset**

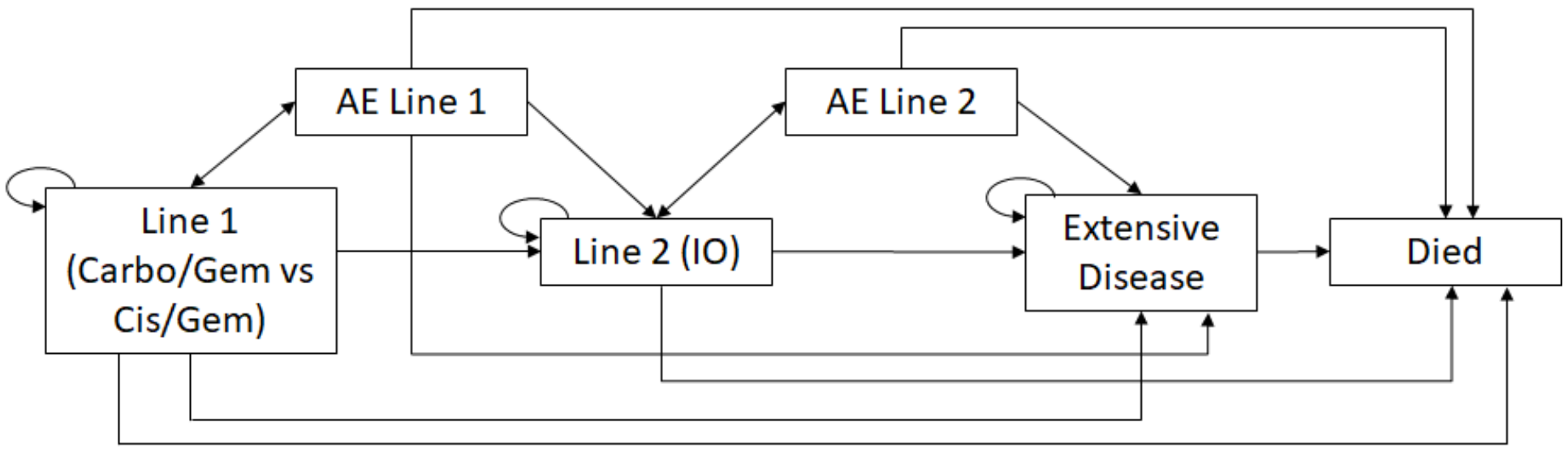

AE= Adverse event
"Extensive Disease" denotes a health state with poor quality of life and high expenses.

**Table 4: Model inputs for bladder treatment-sequence cost-effectiveness microsimulation**

| Parameter | Value | Reference |
|---|---|---|
| **Utilities** | | |
| L1 on therapy | 0.668 | Lin,[27] Taarnhøj[28] |
| L1 Progression-free after therapy | 0.718 | Lin[27] |
| L1 Adverse event | 0.627 | Patterson[29] |
| L2 | 0.60 | Sarfay[30] |
| L2 Adverse event | 0.559 | Patterson[29] |
| ED | 0.52 | Sarfay[30] |
| **AEs - probability of any grade III/IV** | | |
| Carbo/Gem Pr AE | 0.691 | Dogliotti [31] |
| Cis/Gem Pr AE | 0.60 | Dogliotti [31] |
| IO PR AE | 0.15 | Bellmut[32] |
| **Costs** | | |
| Carbo cost per month | 228.69 | CMS[33] |
| Cis cost per month | 223.03 | CMS[33] |
| Gem cost per month | 448.89 | CMS[33] |
| IO (Pembro) cost per month | 15454.08 | CMS[33] |
| L1 (chemo) to treat AE | 10916 | Hale[34] |
| IO cost to treat AE | 8497 | Hale[34] |
| disease management (L1 and L2) | 2,516 | Slater[35] |
| disease management (ED) | 5835.58 | Slater[35] |

**Figure 4: Calibration of model-based overall survival curve to Inverse-Probability Treatment Weighted overall survival curve in Flatiron Health bladder dataset**

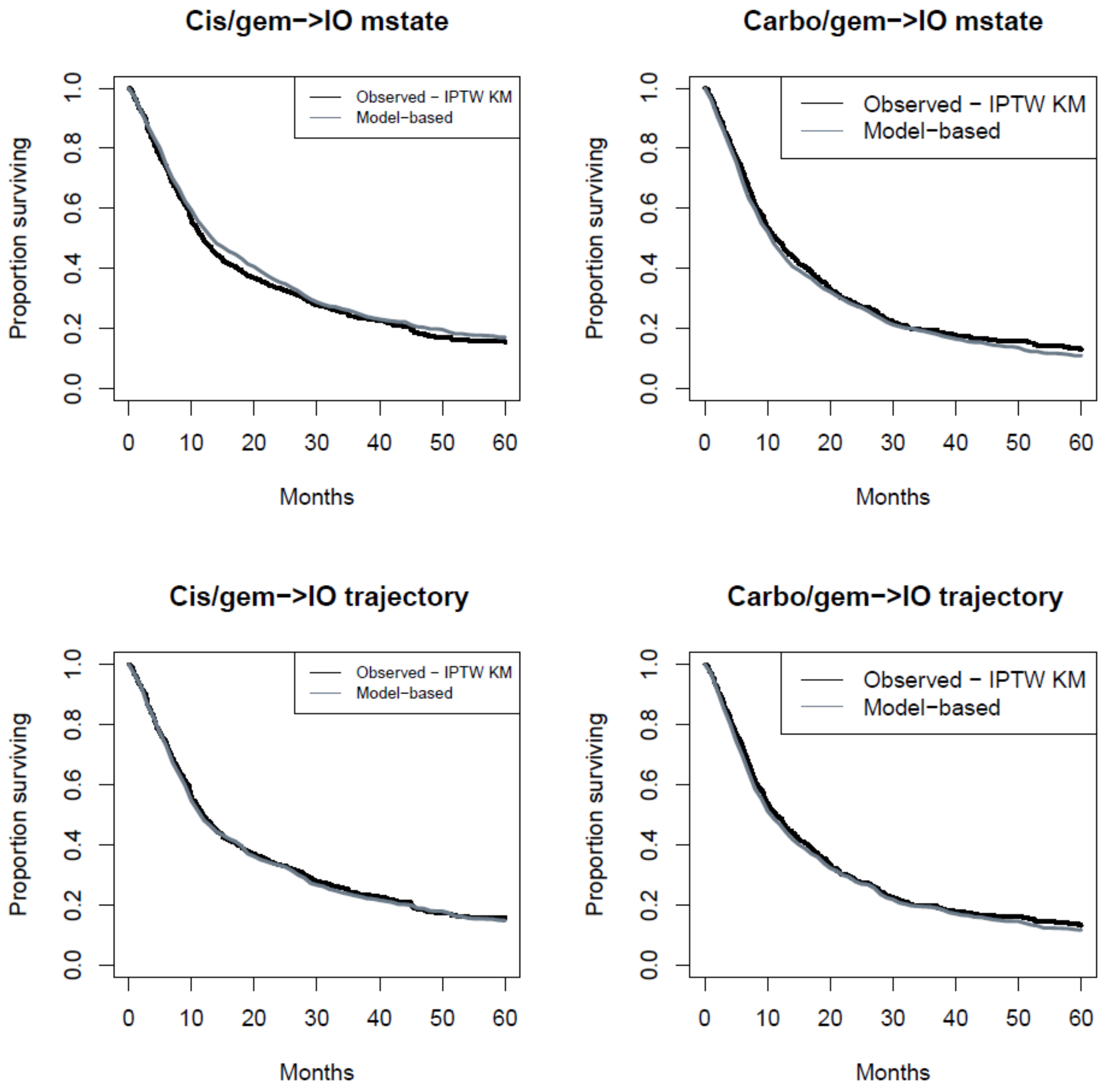

**Table 5: Results of cost/effectiveness analysis of two treatment strategies in Flatiron Health bladder data**

| | | Multi-state models | | | | Trajectory | | | |
|---|---|---|---|---|---|---|---|---|---|
| | | Estimate | SD | 95% CI | | Estimate | SD | 95% CI | |
| Cost | carbo/gem → IO | $127,244 | $1,885 | $123,550 | $130,938 | $125,097 | $4,862 | $115,567 | $134,627 |
| | cis/gem → IO | $138,174 | $2,296 | $133,674 | $142,674 | $126,314 | $5,896 | $114,757 | $137,870 |
| QALY | carbo/gem → IO | 0.954 | 0.017 | 0.920 | 0.988 | 0.967 | 0.041 | 0.887 | 1.047 |
| | cis/gem → IO | 1.175 | 0.019 | 1.138 | 1.212 | 1.105 | 0.046 | 1.014 | 1.196 |
| Diff | Delta Cost | $10,930 | $2,260 | $6,500 | $15,360 | $1,217 | $7,219 | -$12,933 | $15,366 |
| | Delta QALY | 0.221 | 0.016 | 0.189 | 0.253 | 0.138 | 0.054 | 0.033 | 0.243 |
| NMB | WTP=100,000 | $11,185 | $1,892 | $7,477 | $14,893 | $12,572 | $4,612 | $3,532 | $21,611 |

We find excellent calibration between the model-based overall survival curves and weighted observed overall survival. Formally, testing for lack-of-fit using the Kruskal-Wallis test, none of the model-based curves demonstrated lack of fit (P>0.05). Fit is slightly better in the trajectory case, with lower sum-of-squared-errors. We note that the exponential simplification did result in lack of fit, so these models re-estimated transition probabilities at each cycle.

Both modeling approaches showed larger QALYs and larger costs for the cis/gem→IO strategy compared to the carbo/gem→IO, although cost differences were modest. Differences in QALYs were larger using the mstate approach than the trajectory approach (0.221 vs 0.135, respectively), however, both were significantly greater than zero, which is consistent with clinical literature.[31] Higher costs were significant using the mstate approach, but not the trajectory approach. Note that drug costs for cisplatin and carboplatin are very similar; the difference in costs is mainly driven by monthly costs of caring for advanced cancer patients. Both methods showed a Net Monetary Benefit >0 at a willingness to pay of 100,000 per QALY.

## 5. Discussion

We have demonstrated two novel microsimulation models which leverage longitudinal data collected from the EHR to conduct cost-effectiveness analysis. We showed that our approach produces results which are well calibrated to the observed survival data, even though this is not modeled directly. Using observed patient trajectories until censoring was most successful and should be considered for model-based studies using EHR data.

One of the key strengths of this approach is that it allows incorporation of parameters that are not measured in the medical record. Billing data is not usually integrated into the EHR and is often challenging to obtain for direct study. Quality of life measures are also not routinely taken, especially longitudinally as patient's health status changes. Finally, adverse events are difficult to obtain reliably from the EHR, and outside of a trial they are unlikely to be graded.[36,37] The approach we develop allows researchers to incorporate external data into a model while still retaining the benefits of a real-world patient cohort. Although this approach is motivated by studying therapy sequence, the general framework could be used for a variety of clinical and cost-effectiveness questions.

There are a few drawbacks of our approaches. First, we use discrete time models, which do not always perform as well as continuous time models.[38] Second, because of the use of multi-state models with bootstrap resampling, this approach is quite computationally intensive. Finally, there was some bias in the estimates when comparing microsimulation-based results to true values in the synthetic data sets. Although our method substantially outperformed a Markov cohort approach, further refinements are needed.

By creating a synthetic dataset with several different copula models, we investigated several different types of within-patient dependence. Nevertheless, we found that a large amount of the within-patient dependence was driven by covariate effects. Even in the independence case, there was notable correlation in outcomes over time, and the standard Markov cohort approach underestimated effect outcomes.

A key feature of the multi-state model is its ability to simultaneously estimate transition probabilities to different states, conditional on covariates and time. We had also considered competing risks regressions using Fine and Gray models,[39] however, these models have a drawback: for certain covariate patterns, the cumulative incidence of multiple types of events can exceed 1.[40] There are alternative cause-specific models;[41] using these will be an area of future research.

This work has several important limitations. First, as our example was intended to be demonstrative of the method, we did not conduct sensitivity analyses for our parameter estimates of cost, utility, or adverse event rates. We also simplified all adverse events into a single composite. In a full clinical application, the adverse events should be considered more thoroughly. Another drawback is that the mstate approach here used a Cox model.[19] Our log-normal and log-logistic scenarios violate the proportional hazards assumption; unsurprisingly, calibration was worse in these cases. We discuss measures of uncertainty here, and although we computed confidence intervals for our estimates, we did not formally assess coverage. Finally, we assume that progression times can be obtained from the EHR. Although this is available in the Flatiron Health dataset, this is based on abstraction,[26] which may not always be available. Algorithmic identification of progression has been proposed, but it continues to be a challenge for EHR-based studies.[42]

Despite these limitations, this work presents a novel method for using longitudinal EHR data to inform microsimulation models for cost-effectiveness analysis. Future directions in this area should include investigating cause-specific hazards models, relaxing the assumption of proportional hazards, and improving computational efficiency.

Supplemental tables and figures

Table S1: Logistic model parameters, where resulting probabilities were used to generate synthetic treatment sequence, and whether a patient received L2 or went directly to ED

| | Treatment sequence model | Receive L2 therapy after progression |
|---|---|---|
| (Intercept) | -10.600 | -2.349 |
| Seq 1 | NA | 0.193 |
| Surgery | 0.087 | 0.372 |
| Male | -0.392 | -0.290 |
| Race Black | -0.222 | 0.001 |
| Race other | -0.118 | -0.021 |
| Hispanic | -0.326 | -0.082 |
| Age | 0.239 | 0.069 |
| AGE^2 | -0.002 | -0.001 |
| ECOG 1 | -0.219 | -0.174 |
| ECOG 2 | -0.380 | -0.561 |
| GFR | 0.080 | -0.010 |
| GFR^2 | -0.00035 | 0.00004 |
| hemoglobin | 0.066 | 0.018 |
| albumin | 0.028 | 0.044 |
| CHF known | -0.709 | 0.605 |
| hearing loss known | 1.306 | -0.175 |
| neuropathy known | -0.379 | -0.112 |
| viscmet known | -0.422 | -0.022 |

Table S2: Survival model parameters used to generate potential transition times in copula models

| **Weibull** | ProgL1 | DeathL1 | ProgL2 | DeathL2 | DeathED |
|---|---|---|---|---|---|
| shape | 0.937 | 0.942 | 0.968 | 1.171 | 0.931 |
| scale | 3.201 | 4.802 | 6.463 | 1.846 | 0.942 |
| seq 1 | 0.356 | 0.259 | -0.007 | 0.208 | -0.206 |
| Surgery | 0.420 | 0.674 | 0.039 | 0.427 | 0.243 |
| Age | 0.003 | -0.019 | 0.014 | -0.003 | -0.015 |
| ECOG 1 | -0.114 | -0.286 | -0.068 | -0.244 | -0.977 |
| ECOG 2 | -0.327 | -1.285 | -0.114 | -0.706 | -1.055 |
| albumin | 0.021 | 0.095 | -0.019 | 0.060 | 0.093 |

| **Log logistic** | ProgL1 | DeathL1 | ProgL2 | DeathL2 | DeathED |
|---|---|---|---|---|---|
| shape | 1.556 | 1.154 | 1.567 | 1.365 | 1.673 |
| scale | 1.998 | 1.413 | 0.860 | 1.461 | 0.598 |
| seq 1 | 0.270 | 0.223 | 0.00004 | 0.169 | -0.235 |
| Surgery | 0.239 | 0.626 | 0.035 | 0.430 | 0.111 |
| Age | 0.003 | -0.019 | 0.010 | -0.005 | -0.007 |
| ECOG 1 | -0.125 | -0.280 | -0.061 | -0.310 | -0.547 |
| ECOG 2 | -0.215 | -1.154 | 0.178 | -0.680 | -0.571 |
| albumin | 0.022 | 0.115 | -0.002 | 0.060 | 0.068 |
| **Log normal** | ProgL1 | DeathL1 | ProgL2 | DeathL2 | DeathED |
| meanlog | 0.733 | 0.327 | 0.592 | 0.313 | -0.406 |
| sdlog | 1.164 | 1.618 | 1.093 | 1.215 | 1.056 |
| seq 1 | 0.297 | 0.216 | 0.046 | 0.150 | -0.191 |
| Surgery | 0.267 | 0.576 | 0.059 | 0.442 | 0.136 |
| Age | 0.003 | -0.017 | 0.013 | -0.005 | -0.009 |
| ECOG 1 | -0.121 | -0.239 | 0.003 | -0.269 | -0.592 |
| ECOG 2 | -0.219 | -1.114 | 0.156 | -0.695 | -0.623 |
| albumin | 0.021 | 0.114 | -0.001 | 0.063 | 0.071 |